\documentclass[twocolumn]{aastex631}
\usepackage{xspace}

\newcommand{\gammaray}{$\gamma$-ray\xspace}
\newcommand{\gammarays}{$\gamma$-rays\xspace}
\newcommand{\gmgm}{$\gamma$--$\gamma$\xspace}
\newcommand{\Fermilat}{{\it Fermi}-LAT\xspace}

\begin{document}

\title{Searching for Internal Absorption Signatures in High-Redshift Blazars}

\author[0000-0003-0102-5579]{A.~Dmytriiev}
\affiliation{Centre for Space Research\\
North-West University\\
Potchefstroom, 2520, South Africa}
\email{anton.dmytriiev@nwu.ac.za}

\author[0000-0002-2028-9230]{A.~Acharyya}
\affiliation{CP3-Origins, University of Southern Denmark, Campusvej 55, 5230 Odense M, Denmark}
\email{atreya@cp3.sdu.dk}

\author{M.~B\"{o}ttcher}
\affiliation{Centre for Space Research\\
North-West University\\
Potchefstroom, 2520, South Africa}

\begin{abstract}

The \gammaray emission from Flat Spectrum Radio Quasars (FSRQs), a sub-class of blazars, is believed to be generated through interactions of high-energy leptons and/or hadrons in the jet with the ambient photon fields, including those from the accretion disk, the broad line region (BLR), and the dusty torus. However, these same photon fields can also attenuate \gammarays through internal photon-photon (\gmgm) absorption, imprinting characteristic spectral features. Investigating the internal absorption is crucial for unraveling the complex structure of FSRQs and constraining the poorly known location of the \gammaray emission region. In this study, we select a sample of \gammaray detected FSRQs with high redshift ($z \gtrsim 3$), to search for absorption features appearing at lower photon energies due to a substantial redshift. We extract the \Fermilat \gammaray spectra of these sources and perform physical modeling using a detailed \gmgm opacity model, assuming that the BLR photon field dominates the absorption and focusing on the energy range $\sim 25 \ \text{GeV}/(1+z)$, where the absorption feature due to Ly$\alpha$ photons is expected. Our analysis reveals a hint of internal absorption for one source (the lowest redshift object in our sample, $z \simeq 3$) and provides constraints on the location of its \gammaray emitting region along the jet. For the remaining, higher-redshift sources, the limited photon statistics prevent a reliable detection of internal opacity features.

\end{abstract}


\section{Introduction} \label{sec:intro}

Blazars are a  subclass of Active Galactic Nuclei (AGN), with jets that happen to point very close towards the Earth. The spectral energy distribution (SED) of blazars represents a typical two-bump feature \citep[e.g.,][]{fossati1998, ghisellini2017}. The low-energy bump is commonly attributed to synchrotron emission from relativistic electrons within the jet. However, the origin of the high-energy bump remains under debate. In leptonic models, this high-energy emission is attributed to inverse Compton (IC) scattering, where energetic leptons in the jet interact with a field of lower-energy photons. These photons may either be the synchrotron photons themselves (in the synchrotron self-Compton (SSC) process) \citep[e.g.,][]{maraschi1992, bloommarscher1996}, or come from an external source, such as the broad-line region (BLR) \citep[e.g.,][]{sikora1994}, the accretion disk \citep[e.g.,][]{dermer1992}, or the dusty torus \citep[e.g.,][]{blazejowski2000} (in the external inverse Compton (EIC) process). In hadronic scenarios, the high-energy photons are produced in proton-photon interactions through the decay of neutral pions or through proton synchrotron emission \citep[e.g.,][]{mannhbier1992, mannheim1993, mucke2003, boettcher2013}. Blazars are further divided into two categories, BL Lacertae (BL Lac) objects and Flat Spectrum Radio Quasars (FSRQs). Within the leptonic framework, the \gammaray emission of BL Lac objects is typically well explained by the SSC mechanism, whereas for FSRQs, the EIC scenario is favored \citep[e.g.,][]{ghisellini2010, boettcher2013}.

The specific target photon field for \gammaray production (whether leptonic or hadronic origin) strongly depends on the distance of the emission zone from the central supermassive black hole (SMBH), $d$. In order of increasing distance, the photon field is initially dominated by the accretion disk close to the central engine, followed by the broad-line region (BLR) as the primary contributor, and ultimately by the dusty torus at the largest distances. The exact location of the blazar \gammaray emitting region along the jet remains poorly constrained, with studies proposing a range of possibilities. Some studies suggest an emitting region near the outer edge of the BLR \citep[e.g.,][]{dermer2014, paliya2015, boettcherels2016}, while others point to locations beyond the BLR \citep[e.g.,][]{costamante2018, hess2019}, or even much farther away, likely in the vicinity of the dusty torus \citep[e.g.,][]{tavecchio2013}. Additionally, evidence from some studies supports a more dynamic and complex scenario, involving an emitting region that moves rapidly along the jet \citep[e.g.,][]{hayashida2012, dmytriiev2023}, or multiple emitting regions within the jet \citep[e.g.,][]{acharyya2021}.

One powerful method to constrain the location of the \gammaray emitting region is through the study of internal \gmgm absorption. The \gammaray beam generated in blazars traverses photon fields produced by the AGN, with \gammaray photons potentially undergoing attenuation via photon-photon pair production. This creates a delicate balance: while the emitting region must interact with a sufficient radiation field to generate \gammarays, the density of this field cannot be excessively high as to cause catastrophic internal \gmgm opacity. Depending on the exact location of the \gammaray emitting zone along the jet, as well as the density of the target photon field, \gmgm absorption may introduce distinct features in \gammaray spectra, which can be detected \citep[e.g.,][]{poutanenstern2010}. In particular, if the \gammaray emitting zone is located close to the BLR \citep[e.g.,][]{ghisellini2010, boettcherels2016}, the dominant soft photons are provided by the UV emission of the BLR, with the most prominent component being the Ly$\alpha$ emission line (rest frame energy of a photon $\simeq 10.2$ eV). The characteristic energy of \gammarays absorbed by interactions with the Ly$\alpha$ photons is given by the pair-production threshold condition $\epsilon_{\rm s} \epsilon_{\gamma} \geq 2 (1-\mu)^{-1}$, with $\epsilon_{\rm s}$ and $\epsilon_{\gamma}$ being the dimensionless (in the units of the electron rest energy) energy of the soft and \gammaray photons, respectively, and $\mu$ being the cosine of the angle between wave vectors of the two photons. For head-on collision, $\epsilon_{\rm s} \epsilon_{\gamma} = 1$ at threshold, which yields $E_{\gamma,\mathrm{obs}} \simeq 25 \ \text{GeV} / (1+z)$, with $z$ being the redshift of the source. If the redshift of the source is significant, the opacity features are shifted to lower energies, where the sensitivity of the \Fermilat is better. The detection of \gammaray emission at these characteristic energies, along with exact level of its flux, enables constraints on the strength of internal opacity in a given source by setting an upper limit on absorption. Conversely, a non-detection in this energy domain does not provide such information, leaving the opacity strength unconstrained.

In this work, we perform the analysis and detailed physical modeling of \Fermilat data from nine high-redshift ($z \gtrsim 3$) blazars. We search for characteristic features in the \gammaray spectra induced by the internal absorption on the BLR photon field, with an aim to derive constraints on the unknown distance of the \gammaray emitting zone from the SMBH along the jet for the sources in our sample.

\section{Data} \label{sec:data}

We select nine $\gamma$-ray-detected high-z blazars using the sample from \cite{paliya2020}, all of which are FSRQs. The target photon field density, and consequently the internal opacity within each source, depends on the accretion disk luminosity, for which we adopted the values from \cite{paliya2020}, derived from the broad-band SED modeling of these blazars. Table~\ref{tab:sample} summarizes the key parameters for our selected sources. For simplicity, each source will henceforth be referred to by its corresponding number in the first column of the table.

\begin{table*}
  \setlength{\tabcolsep}{14pt}
  \centering
    \begin{tabular}{ccccc}
Source \# & Name (NVSS) & Name (4FGL) & $z$ & $\text{log}_{10}$ [$L_{\rm D}$ (erg s$^{-1}$)] \\
\hline
1 & J033755-120404 & J0337.8-1157 & 3.442  &   46.36 \\
2 & J053954-283956 & J0539.9-2839 & 3.104  &   46.70 \\
3 & J073357+045614 & J0733.8+0455 & 3.01   &   46.60 \\
4 & J080518+614423 & J0805.4+6147 & 3.033  &   46.34 \\
5 & J083318-045458 & J0833.4-0458 & 3.5    &   47.15 \\
6 & J135406-020603 & J1354.3-0206 & 3.716  &   46.78 \\
7 & J142921+540611 & J1428.9+5406 & 3.03   &   46.26 \\
8 & J151002+570243 & J1510.1+5702 & 4.313  &   46.63 \\
9 & J163547+362930 & J1635.6+3628 & 3.615  &   46.30 \\
    \end{tabular}
    \caption{Various observational quantities related to the nine sources in our sample. The redshifts and accretion disk luminosities are adopted from \cite{paliya2020}.}
    \label{tab:sample}
\end{table*}

\subsection{\Fermilat}

The \emph{Fermi}-Large Area Telescope (LAT; \citealt{Fermi_LAT}) is a pair conversion telescope capable of detecting \gammaray photons in the energy range from 20~MeV to above 500 GeV. Primarily operating in survey mode, the \Fermilat scans the entire sky every three hours. In this work, we analyzed \textit{Fermi}-LAT data between MJD~54683 and MJD~59794, which corresponds to August 4, 2008, the start of the \Fermilat mission until midnight on August 4, 2022. Throughout the analysis, we use the \textit{Fermi} Science Tools version $11-05-03$\footnote{\url{http://fermi.gsfc.nasa.gov/ssc/data/analysis/software} (accessed on 10/05/2023)} and \texttt{FERMIPY} version 1.0.1~\footnote{\url{http://fermipy.readthedocs.io} (accessed on 10/05/2023)} \citep{wood2017fermipy} in conjunction with the latest \textit{PASS} 8 IRFs~\citep{atwood2013pass}.

\gammaray photons having energies between 100 MeV and 300 GeV that were detected within a region of interest (RoI) of radius 15$^{\circ}$ centered on the location of each source in the sample were selected for the analysis. Furthermore, we selected only photon events from within a maximum zenith angle of 90$^{\circ}$ in order to reduce the contamination from background photons from the Earth's limb, produced by the interactions of cosmic-rays with the upper atmosphere.
The contributions from the isotropic and Galactic diffuse backgrounds were modeled using the most recent templates for isotropic and Galactic diffuse emission, iso\_P8R3\_SOURCE\_V3\_v1.txt and gll\_iem\_v07.fits, respectively. Sources in the 4FGL-DR4 catalog \citep{4fgl_dr4} within a radius of $20^{\circ}$ from the position of each source in the sample were included in the model, with their spectral parameters fixed to the catalog values. This takes into account the \gammaray emission from sources lying outside the RoI which might yet contribute photons to the data, especially at low energies, due to the size of the point spread function of the \Fermilat.

The normalization factors for both the isotropic and Galactic diffuse emission templates were left free, while the spectral parameters for all points sources within the ROI (excluding the source of interest) were fixed to the values reported in the 4FGL-DR4 catalog. A binned likelihood analysis was then performed, with a complex spectral model incorporating the \gmgm opacity (see Section~\ref{sec:liklha_results}).

\subsection{Emission line data}
\label{subsec:emldata}

The features in the \gammaray spectra induced by \gmgm opacity within the source are highly dependent on the shape of the target photon field. In this study, we assume that the BLR photon field is the dominant contributor to both \gammaray production (whether leptonic or hadronic processes) and \gmgm absorption for all selected objects. Unlike previous studies that rely exclusively on average characteristic templates of the BLR spectrum or assume constant fractions of the accretion disk luminosity for the luminosity of emission line(s), we employ direct measurements of BLR line luminosities (where available), complementing them with reliable scaling relations based on well-established average emission line luminosity ratios. This hybrid approach ensures a more robust and self-consistent estimation of the target photon field and internal opacity. The emission lines of the BLR typically represent a substantial fraction of its radiative output, which implies that the energy-dependent opacity is highly sensitive to the fluxes of these individual lines. Due to the high redshifts of our sources, source-frame UV lines are observed in the optical -- IR bands on Earth.

Of particular importance is the accurate determination of the Ly$\alpha$ emission line (rest-frame 1216 \AA), one of the most prominent lines. This line induces opacity features in the \gammaray spectra at the lowest energies. Since the Ly$\alpha$ line overlaps with the neighboring N~V line (rest-frame 1240 \AA), we do not attempt to separate them in our modeling. Instead, we use the combined Ly$\alpha$+N~V flux, centering the composite line at the Ly$\alpha$ wavelength (rest-frame 1216 \AA), and refer to it simply as the Ly$\alpha$ flux. In addition to Ly$\alpha$, we consider three other prominent emission lines at longer wavelengths: C~IV (rest-frame 1549 \AA), Mg~II (rest-frame 2798 \AA), and H$\beta$ (rest-frame 4861 \AA).

To determine the line luminosities, we first explore archival data sources. The C~IV line luminosities are available from \cite{paliya2021}, except for Source 3. These measurements can be considered reliable due to the robust methodology employed by the authors. The Mg~II and H$\beta$ line luminosities are sourced from the Gemini near-infrared spectroscopy study by \cite{burke2024}, although the quality of spectra varies across our sample, with some lines exhibiting significant noise contamination. This limitation affects our ability to accurately measure Mg~II and H$\beta$ fluxes for certain sources.

Obtaining the Ly$\alpha$ luminosity proves to be the most challenging. Unlike other emission lines, where the continuum can be measured and subtracted on both sides, the bluer side of Ly$\alpha$ is heavily contaminated by the Ly$\alpha$ forest. Additionally, there is very limited available data for Ly$\alpha$. For four sources in our sample, the Ly$\alpha$ flux could be derived from the optical spectra provided by the Sloan Digital Sky Survey (SDSS) Data Release (DR) 18 \citep{sdssdr18}. However, we opted not to rely on these measurements, as the SDSS spectra have not been corrected for Galactic extinction, and applying accurate corrections involves a rather complex procedure.

To address these challenges, we adopt the reliable C~IV line luminosities from \cite{paliya2021}. For the other lines (Ly$\alpha$, MgII, and H$\beta$), we scale the C~IV luminosities using the average relative emission line luminosity ratios provided by \cite{finke2016} (Table 5), which updates the widely used composite quasar spectrum from \cite{francis1991}. This method employs the average luminosity ratios for Ly$\alpha$/Mg~II/H$\beta$ relative to C~IV, as given in \cite{finke2016}, applied to the measured C~IV luminosities for each source. The Ly$\alpha$ ratio already includes the contribution of the N~V line, thus providing the overall flux.

For Source 3, where C~IV line data is unavailable, we use the Mg~II line luminosity measurement from \cite{burke2024}, where the spectrum and emission line quality are high, ensuring reliability of the luminosity estimate. We then apply the Ly$\alpha$/C~IV/H$\beta$ scaling ratios relative to Mg~II from \cite{finke2016} to predict the luminosities of the remaining lines. This scaling approach is expected to yield reasonably accurate results, with a systematic uncertainty in the line ratios of approximately 5\% \citep{finke2016}. However, it is important to note that emission line luminosities can vary in time rather significantly, typically by a factor of 3 -- 4 \citep[e.g.,][]{dmytriiev2023}. Due to the lack of optical data from multiple epochs, we assume that the derived emission line luminosities represent represent time-averaged values over the period covered by the \textit{Fermi}-LAT observations. The final line luminosities for the blazar sample are summarized in Table~\ref{tab:emlines}.

\begin{table*}
  \setlength{\tabcolsep}{10pt}
  \centering
    \begin{tabular}{ccccc}
Source \# & $\text{log}_{10}$ [$L_{\mathrm{Ly}\alpha\mathrm{+NV}}$ (erg s$^{-1}$)] & $\text{log}_{10}$ [$L_{\mathrm{C~IV}}$ (erg s$^{-1}$)] & $\text{log}_{10}$ [$L_{\mathrm{Mg~II}}$ (erg s$^{-1}$)] & $\text{log}_{10}$ [$L_{\mathrm{H}\beta}$ (erg s$^{-1}$)]  \\
\hline
1   & 44.885  &  44.268 $\pm$ 0.193  &  44.036  &  43.806  \\     
2   & 45.708  &  45.091 $\pm$ 0.091  &  44.859  &  44.629  \\       
3   & 44.94   &  44.32  $\pm$ 0.02   &  44.09   &  43.86   \\         
4   & 45.36   &  44.743 $\pm$ 0.095  &  44.511  &  44.281  \\       
5   & 45.837  &  45.220 $\pm$ 0.111  &  44.988  &  44.758  \\       
6   & 45.169  &  44.552 $\pm$ 0.038  &  44.320  &  44.09   \\       
7   & 44.858  &  44.241 $\pm$ 0.018  &  44.0    &  43.779  \\        
8   & 45.474  &  44.857 $\pm$ 0.059  &  44.625  &  44.395  \\       
9   & 44.922  &  44.305 $\pm$ 0.023  &  44.073  &  43.843  \\       
\end{tabular}
    \caption{Measured and derived luminosities of the four dominant emission lines in the BLR spectra of the blazar sample, used in our modeling to calculate the \gmgm opacity. The uncertainties for the logarithm of the C~IV luminosity are taken from \cite{paliya2021} (except for Source 3). Due to the scaling procedure employed, the uncertainties for the logarithms of all other emission line luminosities are identical to those for C~IV. For Source 3, the uncertainty is used for the logarithm of the Mg~II luminosity \citep{burke2024}, and this value is the same for the other lines for this source.}
    \label{tab:emlines}
\end{table*}

\section{Internal opacity modeling} \label{sec:iom}

We use the \gmgm opacity code by \cite{boettcherels2016}, which performs a full angle integration for the interaction of a \gammaray with incident soft photons from the BLR, and employs a fully angle-dependent \gmgm interaction cross-section. The BLR in this model is assumed to have a shell-like geometry with inner and outer radii $R_1$ and $R_2$, respectively. The code performs normalization of the BLR emissivity, via full angle integration, to match the correct value of the BLR field density inside the BLR (given as a parameter). The luminosity of the BLR is a fixed quantity, while the energy density of the BLR photon field varies with distance $d$. The code computes the optical depth $\tau_{\gamma-\gamma}$ as a function of \gammaray energy $E_{\gamma}$ and the distance of the emitting zone from the central engine, $d$. Following \cite{boettcherels2016}, we assume $R_1 = 0.9 \ R_{\rm BLR}$, $R_2 = 1.1 \ R_{\rm BLR}$, with the exact choice of the values having a negligible impact on the final results.

As the template of the target photon field for the \gmgm absorption, $u_{\nu}(\nu)$, used by the code, we use the four above-mentioned most dominant emission lines superimposed on a blackbody continuum with a fixed temperature $T=1500$ K. The inclusion of a continuum component, in addition to the emission lines, is motivated by the aim to construct a physically realistic representation of the BLR radiation field, as observed in quasar spectra. The BLR continuum is generally well described by a blackbody spectrum with an effective temperature of $T \sim 1000$ K \citep[e.g.,][]{czerny2011}, but a higher temperature of 1500 K may be more appropriate when accounting for contributions from molecular clouds or dust sublimation regions \citep[e.g.,][]{baskinlaor2018}.

For the emission lines, we use the previously obtained information on the luminosity of these emission lines in the BLR spectrum (see Table~\ref{tab:emlines}). The energy density of a given emission line within the source is estimated as

\begin{equation}
    u_{\rm line} = \frac{L_{\rm line}}{4\pi R_{\rm BLR}^2 c}
\end{equation}
where $L_{\rm line}$ is the line luminosity, $R_{\rm BLR}$ is the BLR size, and $c$ is the speed of light in vacuum. We estimate the size of the BLR based on the accretion disk luminosity using the standard scaling relation $R_{\rm BLR} \simeq 0.1 (L_{\mathrm{D},46})^{1/2}$ pc, with $L_{\mathrm{D},46} = L_{\rm D} / (10^{46} \ \text{erg} \ \text{s}^{-1})$ \citep[Eq.~2 in][]{baskinlaor2018}.

To maintain consistency across the sample, we assume that the total luminosity of the BLR, $L_{\rm BLR,tot}$, is a fixed fraction, $\xi_{\rm BLR}$, of the accretion disk luminosity, $L_{\rm D}$, meaning that the fraction $\xi_{\rm BLR}$ of the disk radiation is reprocessed in the BLR. This results in a total BLR energy density of

\begin{equation}
    u_{\rm BLR,tot} = \frac{\xi_{\rm BLR} L_{\rm D}}{4\pi R_{\rm BLR}^2 c}
\end{equation}

Typical values of the BLR covering factor, $\xi_{\rm BLR}$, are uncertain, with estimates ranging from 5 -- 10\% for radio-quiet AGN \citep[e.g.,][]{oke1982} and up to 30\% in some cases \citep[e.g.,][]{maiolino2001}. For blazars, measurements are more challenging due to jet emission outshining the BLR, with estimates of $\xi_{\rm BLR}$ around 4 -- 10\% \citep[e.g.,][]{delia2003, zhang2015}. In our sample, the highest ratio of the combined luminosity of the four emission lines to the disk luminosity, $\sum_{1}^{4} L_{\rm line} / L_{\rm D}$, is $\approx 15.4$\%, necessitating a slightly larger covering factor. We therefore adopt $\xi_{\rm BLR} = 20$\% as the smallest multiple of 10 that ensures the total BLR luminosity, $\xi_{\rm BLR} L_{\rm D}$, exceeds the sum of the emission line luminosities for all sources, and thus prevents unphysical negative continuum energy densities.

Given the scaling of $R_{\rm BLR}$ with $L_{\rm D}$, and the fixed assumed fraction $\xi_{\rm BLR} = 0.2$, the total energy density of the BLR remains at a constant value of $u_{\rm BLR,tot} \approx 0.056$ erg cm$^{-3}$ for all sources. The energy density of the blackbody continuum (the normalization) of the utilized BLR spectrum template is therefore calculated as the difference between the total BLR energy density and the sum of energy densities of the four emission lines:

\begin{equation}
    u_{\rm cont} = u_{\rm BLR,tot} - \sum_{1}^{4} u_{\rm line}
\end{equation}

We find that the specific choice of $\xi_{\rm BLR}$, as well as the exact temperature of the blackbody continuum, have only a minor impact on the final results, since the narrow emission lines dominate the \gmgm opacity. At the same time, while the contribution of continuum to \gmgm opacity is subdominant, its inclusion ensures consistency with realistic BLR spectra and maintains the total BLR energy density in line with the assumed 20\% reprocessing fraction.

\begin{figure*}[t]
    \centering
    \begin{minipage}[t]{0.44\textwidth}
        \centering
        \includegraphics[width=\textwidth]{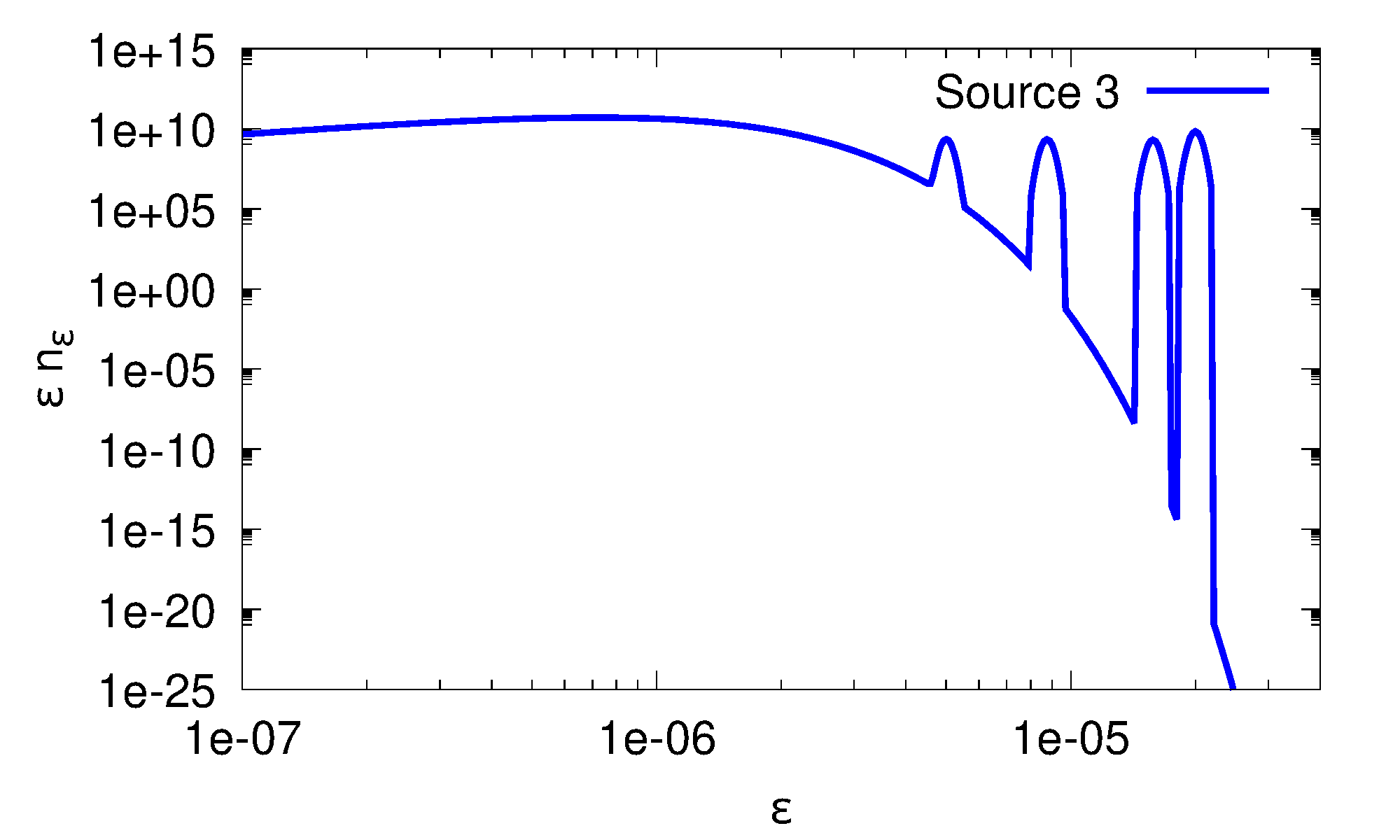} \\
        (a)
    \end{minipage}
    \hfill
    \begin{minipage}[t]{0.54\textwidth}
        \centering
       \includegraphics[width=\textwidth]{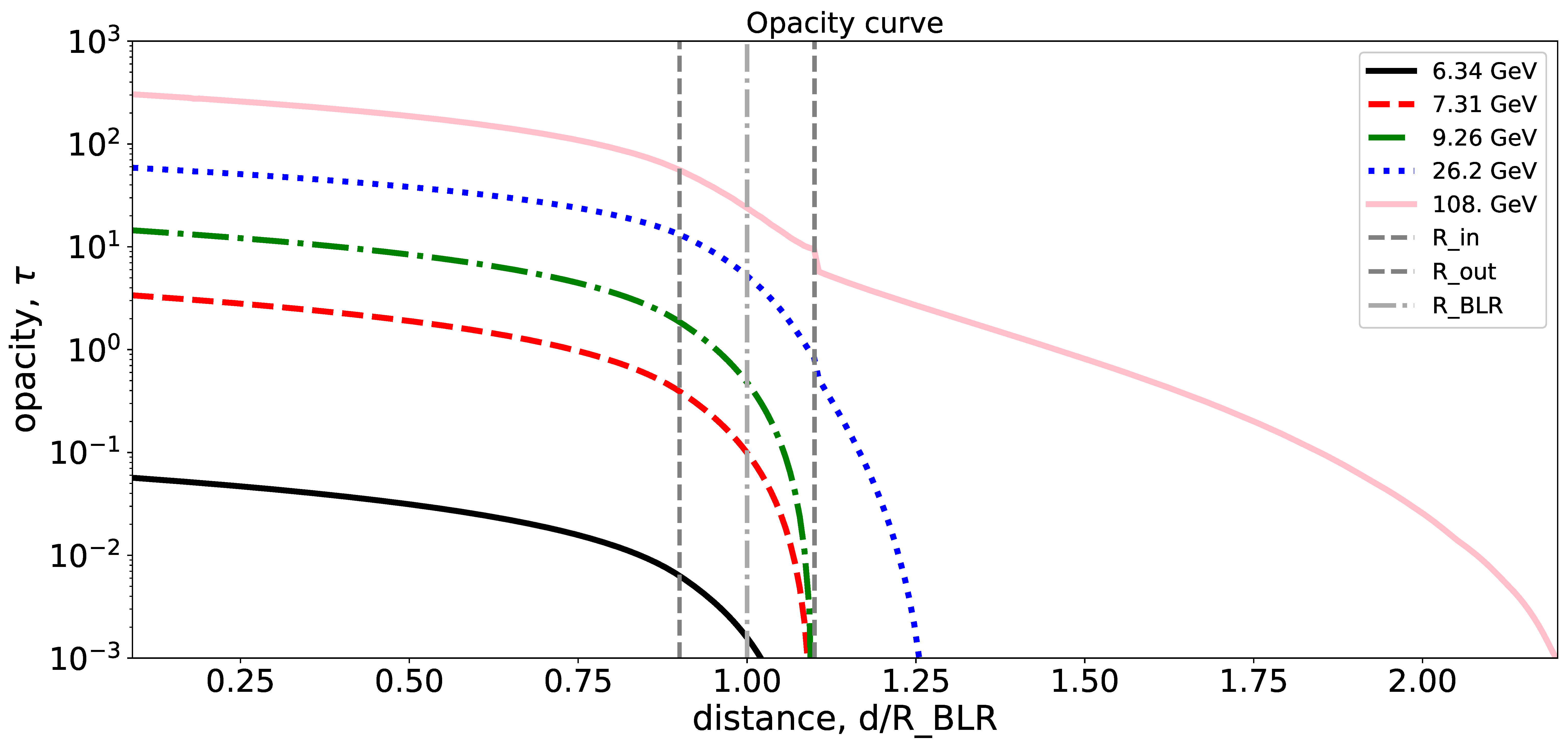} \\
        (b)
    \end{minipage}
    \caption{(a) Target photon field template for the Source 3, shown in the $\epsilon n_{\epsilon}$ representation, with $\epsilon = h\nu/m_{\rm e} c^2$ being the dimensionless target photon energy, and $n_{\epsilon}$ being the number density of target photons per unit of dimensionless energy interval. (b) Optical depth of internal \gmgm absorption $\tau_{\gamma-\gamma}$ depending on the distance $d$, computed for Source 3 for five different \gammaray energies. Here, $R_{\rm in} = R_1 = 0.9 \ R_{\rm BLR}$, $R_{\rm out} = R_2 = 1.1 \ R_{\rm BLR}$.}
        \label{fig:templ_tau_example}
\end{figure*}

An example of the target photon field template is depicted in Fig.~\ref{fig:templ_tau_example}(a). As a result, based on the computed templates, we calculate the 2D table of attenuation factors exp$(-\tau_{\gamma-\gamma})$ on a fine grid of \gammaray energies $E_{\gamma}$ and the distances $d$, for all nine sources in our sample. To visualize the obtained results, an example opacity curve, i.e.\ optical depth $\tau_{\gamma-\gamma}$ as a function of distance $d$ for a number of different \gammaray energies is presented in Fig.~\ref{fig:templ_tau_example}(b). It is worth to note that the optical depth (for the same \gammaray energy) varies greatly among the sources in our sample.

\section{Likelihood analysis and Results} \label{sec:liklha_results}

In our binned likelihood analysis, we use a composite spectral model for each source, comprising three multiplicative components:
\begin{equation}
    \frac{dN}{dE}=\frac{dN_{\text{intrinsic}}}{dE} \times \exp {(- \tau_{\gamma-\gamma})} \times \exp{(- \tau_{\text{EBL}})}
\end{equation}

The first component represents the intrinsic \gammaray spectrum before any absorption effects, which we assume follows one of three possible forms: a power law (PL), a log-parabola (LP), or a power law with an exponential cutoff (PLEC). As the initial spectral parameters of these three models (normalization, spectral index, curvature and cutoff energy), we adopt the values from the 4FGL-DR4 catalog. The second component accounts for the internal \gmgm attenuation within the source, calculated according to the methods detailed in Section~\ref{sec:iom}. The third component models the \gammaray absorption due to interactions with the extragalactic background light (EBL) during the propagation of the \gammarays through the Universe, using the model from \cite{Saldana_lopez} and the redshift values given in Table~\ref{tab:sample}. The final composite spectral model, formed by multiplying these three components, is implemented in \texttt{FERMIPY} as a spectral FileFunction\footnote{\url{https://fermi.gsfc.nasa.gov/ssc/data/analysis/scitools/source_models.html\#FileFunction} (accessed on 10/05/2023)}, with the normalization parameter kept free. This model is then used to extract the spectral energy distributions (SEDs) for the nine sources in our sample.

\begin{figure*}
    \centering
    \includegraphics[width=0.89\linewidth]{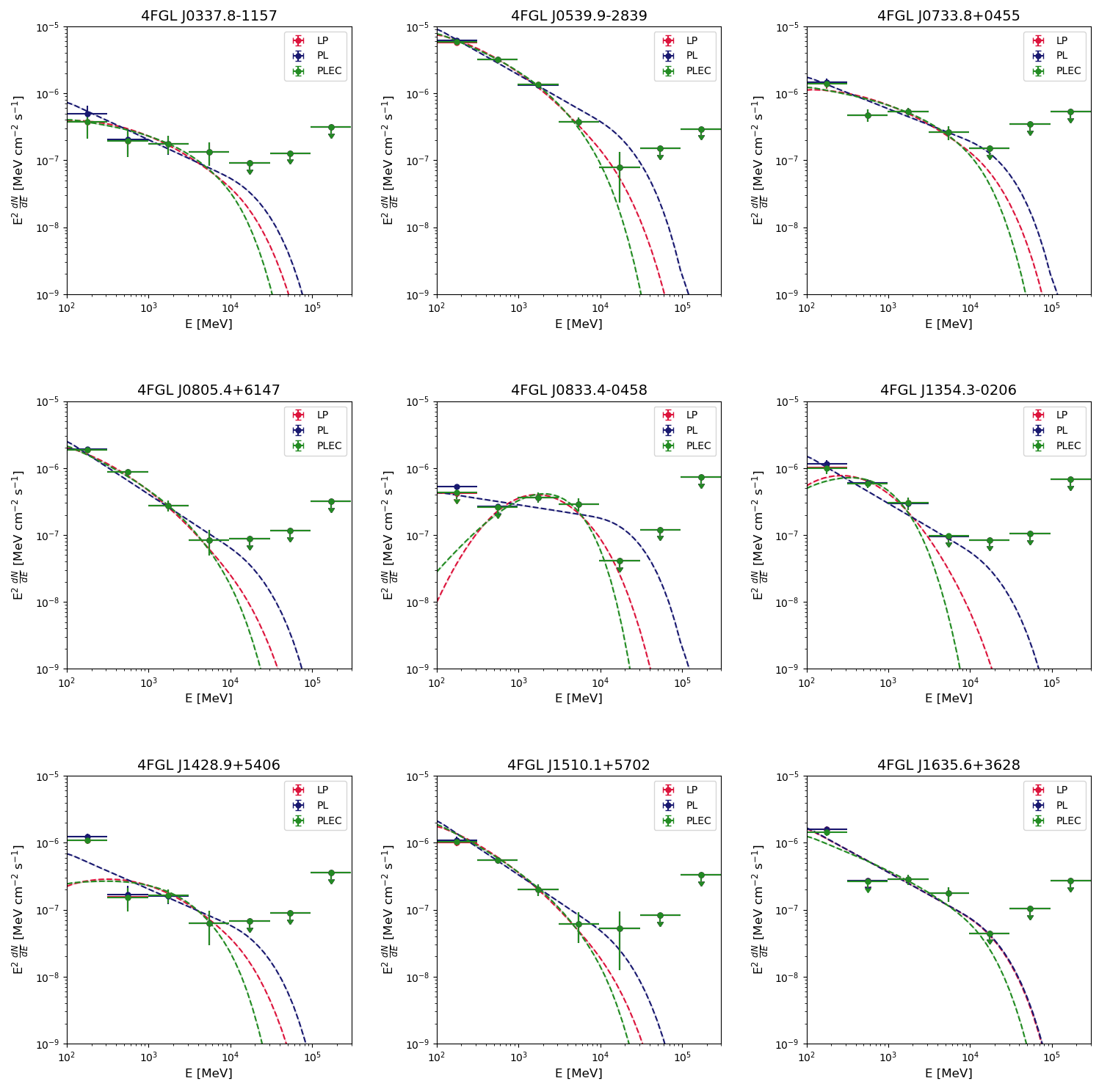}
    \caption{The \Fermilat SEDs for all nine high-z sources in our sample, obtained for the baseline model where the internal \gmgm opacity is set to zero. The data points obtained for each of the  three intrinsic spectral model, PL, LP, and PLEC, are shown in blue, red and green respectively. The corresponding curves represent the best fits obtained for each model. The models account also for EBL absoprtion, using the model from \cite{Saldana_lopez}. The data are binned into two energy bins per decade, with individual bins having a TS $<$ 4 considered as upper limits.}
    \label{fig:spectra}
\end{figure*}

We begin the analysis by extracting the SEDs with a baseline model, shown in Fig~\ref{fig:spectra}, where the internal \gmgm opacity is set to zero (i.e., only the intrinsic PL/LP/PLEC spectrum multiplied by the EBL absorption), and derive the corresponding Test Statistic (TS) for all nine sources, TS$_0$. We have verified that the TS$_0$ values are in a good agreement with the ones in the 4FGL-DR4 catalog. Subsequently, we incorporate the internal opacity into the models and vary the distance $d$ from the SMBH, producing the curve of TS as a function of $d$ for all sources. When varying the distance $d$, only the normalization is left free during likelihood optimization. Allowing other parameters (e.g., spectral index, curvature, etc) to vary would require extensive spectral template generation, while offering only a marginal improvement. This is because internal absorption effects mainly influence the high-energy range (above 5 -- 6 GeV) with limited photon statistics, while the spectral parameters of the 4FGL catalog are likely to be dominated by \Fermilat data at lower energies (with high photon statistics), where the absorption is negligible. Thus, we consider the 4FGL spectral parameters to provide a robust baseline, and allowing only the normalization to vary is expected to be sufficient for capturing the key impacts of internal opacity while maintaining model consistency.

Two scenarios arise from this analysis. If the TS improves when including the internal opacity, it suggests a potential signature of \gmgm absorption. In this case, we determine the distance $d$ that maximizes the TS, calculate its $1\sigma$ uncertainty, and assess the significance of this model using the TS difference between the best-fit model and the baseline model without internal opacity, $\Delta$ TS = TS$_{\rm bf}$-TS$_0$. 

On the other hand, if the inclusion of internal opacity worsens the TS, we derive lower limits on $d$ based on the distance at which the TS value reaches the values corresponding to specific confidence levels. Below this lower limit, the \gmgm absorption predicted by the model becomes inconsistent with the observed \Fermilat photon distribution at the given confidence level. In general, the significance of TS variations is quantified by the TS difference, with $\Delta$~TS~=~4 chosen as the relevant threshold, corresponding to 2$\sigma$ confidence level. The likelihood optimization and TS curve calculation is repeated for all three intrinsic spectral models (PL, LP, and PLEC) to evaluate the impact of the model choice on the final results.

\begin{figure*}
    \centering
    \includegraphics[width=0.89\linewidth]{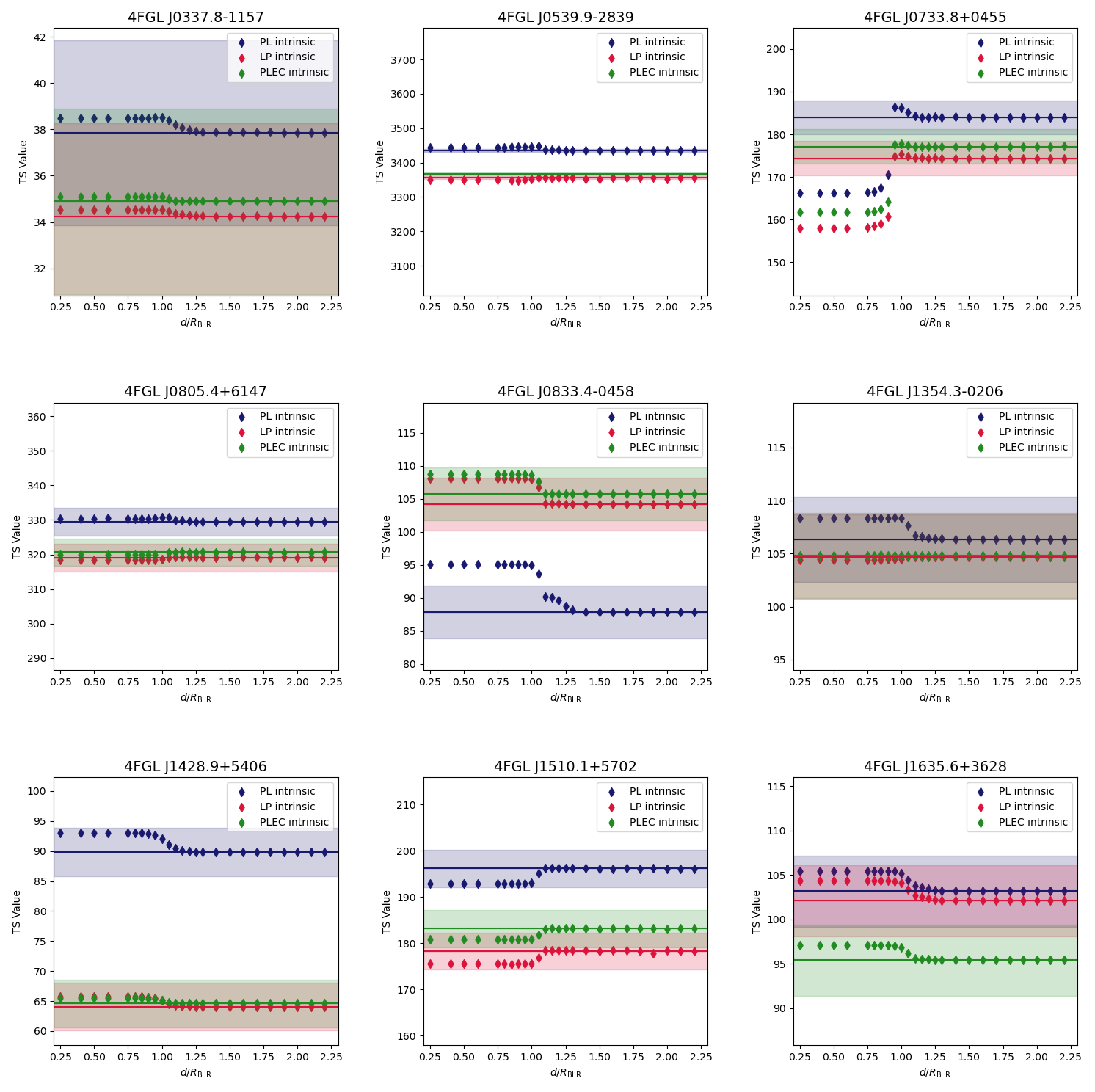}
    \caption{Behavior of TS as a function of the distance from the SMBH $d$ for all nine high-z sources in our sample. Three TS curves are shown, with blue, green and red diamonds representing the power-law, power-law with an exponential cutoff and logparabola intrinsic spectrum models, respectively. The shaded regions visualize the 2$\sigma$ (TS-TS$_0$=4) confidence level.}
    \label{fig:tscurves}
\end{figure*}

The TS curves for all nine sources, calculated for three intrinsic spectral models (PL, LP, and PLEC), are shown in Fig.~\ref{fig:tscurves}. Upon examining the TS behavior, we find that significant changes in TS with distance $d$ ($\Delta$ TS $\geq 4$) are observed only for Source 3 (4FGL J0733.8+0455; z=3.01). For the remaining sources, TS variations remain below this threshold for all intrinsic models, with the exception of the PL model for Source 5 (4FGL J0833.4-0458; z=3.5). However, since the PL model for Source 5 exhibits the lowest TS among the three models, this result can be disregarded. These findings indicate that deviations from the baseline model (which assumes no internal opacity) are not statistically significant for these sources. Consequently, we cannot constrain the distance $d$ for any source except Source 3, for which we find $d \geq 0.88 R_{\rm BLR}$.

\section{Discussion} \label{sec:discuss}

Based on our analysis, we are able to reliably constrain the distance from the SMBH $d$ for only one source in our sample: Source 3 (4FGL J0733.8+0455). This source, having the lowest redshift in the sample (z=3.01), benefits from the highest photon statistics at high energies ($\gtrsim 5$ GeV). In contrast, the remaining sources likely lack sufficient photon statistics in this energy range, preventing similar constraints.

Focusing on Source 3, the TS curve shows an abrupt drop around $d \approx 0.93 R_{\rm BLR}$. Below this distance, the internal opacity model yields a TS difference of $\text{TS} - \text{TS}_0 \approx -18$, compared to the zero-opacity model. This indicates that the observed \Fermilat \gammaray distribution is inconsistent with internal opacity at these distances, with a significance of $\approx 4\sigma$. Thus, the emitting region in Source 3 must be located at $d \geq 0.93 R_{\rm BLR}$, providing a lower limit on $d$. The size of the BLR for Source 3 is $R_{\rm BLR} \approx 0.2$ pc (as derived from the usual scaling relation presented earlier), which implies $d \geq 0.186$ pc, or $d \geq 5.73 \times 10^{17}$ cm.

An intriguing feature is observed for Source 3 in the case of the power-law intrinsic model, just beyond the TS drop. Here, the TS curve exhibits a moderate pile-up and a global maximum, indicating that the opacity model with the emitting region at $d \approx 0.95 R_{\rm BLR}$ provides a better description of the data than the zero-opacity model, quantified by a positive TS difference, $\text{TS}_{\rm bf} - \text{TS}_0 \approx 3$, i.e.\ a confidence level of approximately $1.7\sigma$. Although modest in significance, this suggests a hint of internal \gmgm opacity in Source 3. The exact location of the \gammaray production site can be constrained to $d/R_{\rm BLR} = 0.95^{+0.09}_{-0.02}$, with the derived uncertainties corresponding to a deviation of TS from the best-fit value by 1 ($1\sigma$ level).

The derived lower limit on the distance from the SMBH for Source 3, $d \geq 0.93 R_{\rm BLR}$, suggests that the emitting region is situated at or beyond the inner boundary of the BLR ($d = 0.9 \ R_{\rm BLR}$). It is important to note that, because internal \gmgm opacity increases as $d$ decreases, and due to the ambiguity in interpreting a lack of high-energy photon statistics (which could result from internal \gmgm absorption, EBL absorption, or an intrinsic cutoff in the electron spectrum), an upper limit on the location of the emitting zone along the jet cannot be reliably established. Consequently, only a lower limit can be determined. However, the tighter constraint $d/R_{\rm BLR} = 0.95^{+0.09}_{-0.02}$, equivalent to $0.93 \leq d/R_{\rm BLR} \leq 1.04$, obtained from the analysis of the global TS maximum, indicates a preference for the emitting region to be located within the central portion of the BLR. 

It is important to emphasize that the results obtained in this study are independent of the specific mechanism of \gammaray production, whether leptonic or hadronic. Our analysis focuses solely on the propagation of \gammarays within the source, relying only on the assumption that the (GeV) \gammaray production site is located in the vicinity of the BLR. If, instead, the GeV \gammarays are produced at much farther distances from the SMBH, e.g.\ near the dusty torus, no internal absorption is expected within the \Fermilat energy range. For instance, a study by \cite{costamante2018} of 106 \Fermilat blazars (with relatively low redshifts) found no evidence of internal \gmgm absorption due to interaction with the BLR photons, suggesting that \gammaray production occurs outside the BLR, most likely closer to the dusty torus. Furthermore, while we assume a one-zone emission model, which is a reasonable simplification for majority of blazars, this may not fully reflect the complexity of some sources. Recent studies on \gammaray emitting zone location in FSRQs suggest that (GeV) \gammaray emission can arise from multiple blobs within the BLR \citep[e.g.,][]{finke2016}, or from multiple distinct zones along the jet both near the BLR and the dusty torus \citep[e.g.,][]{acharyya2021}. A more detailed investigation of such multi-zone scenarios, accounting for the propagation effects of \gammarays from multiple emission regions, is beyond the scope of this work and would require a separate analysis. 

In this work, we chose to focus on high-redshift ($z > 3$) sources in which potential BLR absorption features are expected to emerge at relatively low ($E < 6$~GeV) energies, in order to take advantage of {\it Fermi}-LAT's superior sensitivity at $\sim$~GeV energies. However, the downside of this choice is the relative faintness of blazars at such high redshifts, limiting photon statistics. In future work, one might explore sources at intermediate redshifts ($z \sim 1 - 3$) with comparably hard (photon index $\lesssim 2$) $\gamma$-ray spectra. In those, BLR absorption features are expected to emerge at higher energies ($\gtrsim 6 - 12$~GeV), but the expected higher fluxes of such blazars will allow for a more reliable determination of the intrinsic $\gamma$-ray spectra. This, together with the favorable extrapolation of the hard $\gamma$-ray spectra into the $\sim 10$~GeV regime, might allow for more significant constraints on BLR absorption features to be placed than was possible with our high-$z$ sample. An additional, exciting prospect for future studies along these lines will be the advent of the Cherenkov Telescope Array Observatory (CTAO). With the expected low energy threshold ($\sim 10$~GeV) afforded by the Large Sized Telescopes (LSTs), the combination of {\it Fermi}-LAT and CTAO will provide continuous energy coverage of the $\gamma$-ray spectra of $z \sim 1$ blazars from $\sim 100$~MeV to TeV energies. In sources that do not show significant short- to medium-term (days -- weeks) variability (in order to avoid issues with different required integration times to obtain meaningful spectral constraints), this holds great  promise for future, significant detections of BLR absorption features or stringent upper limits on them.

\section{Conclusions} \label{sec:concl}

Our study provides new insights into the location of the \gammaray emitting region in high-redshift blazars by analyzing potential internal \gmgm absorption features. From the nine blazars in our sample, we were able to constrain the distance from the SMBH, $d$, for one source, 4FGL J0733.8+0455 ($z \approx 3.01)$. The results indicate that the \gammaray emitting region in this source is located at $d \geq 0.186$ pc ($d/R_{\rm BLR} \geq 0.93$), with a best-fit value of $d = 0.19^{+0.018}_{-0.004}$ pc ($d/R_{\rm BLR} = 0.95^{+0.09}_{-0.02}$). This suggests the emission originates near the inner boundary or within the central portion of the BLR, consistent with mild internal opacity effects.

For the remaining sources, insufficient photon statistics at high energies limited our ability to impose similar constraints. Our findings emphasize the interplay between photon field densities and internal \gmgm opacity, highlighting the challenges in distinguishing between intrinsic spectral cutoffs and absorption features at these redshifts.

Future studies should target intermediate-redshift blazars ($z \sim 1 - 3$), ideally with harder \gammaray spectra, where BLR absorption features might emerge at higher energies, but with better photon statistics. The upcoming CTAO instrument will complement \Fermilat data, enabling continuous energy coverage and deeper explorations of internal \gmgm opacity. Such advancements promise more precise constraints on the \gammaray emitting regions and the physical environments of blazars.

\begin{acknowledgments}
We thank Paul Els for providing his code for gamma-gamma opacity calculations. We are also grateful to Vaidehi Paliya, Catherine Boisson and Andrew Chen for their insightful discussions and valuable suggestions, which significantly enhanced this work. AD and MB acknowledge support from the Department of Science and Innovation and the National Research Foundation of South Africa through the South African Gamma-Ray Astronomy Programme (SA-GAMMA). AA acknowledges support from Manuel Meyer for guidance and mentorship.
\end{acknowledgments}

%

\vspace{5mm}
\facilities{\Fermilat}


\software{Fermipy version 1.0.1
          }




\bibliography{bibl_highz}{}
\bibliographystyle{aasjournal}



\end{document}